\documentclass[aps,prl,groupedaddress, twocolumn, superscriptaddress, longbibliography]{revtex4-1} 

\usepackage{graphicx}  % needed for figures
\usepackage{amssymb}   % for math
\usepackage{amsmath}
\usepackage{upgreek}
\usepackage{float}
\usepackage[colorlinks=true, pdfstartview=FitV, linkcolor=blue, citecolor=blue, urlcolor=blue]{hyperref} % enable links
\usepackage[free-standing-units=true, per-mode=symbol-or-fraction, range-phrase={-}, separate-uncertainty=true]{siunitx} % for consistent handling of SI units
\DeclareSIUnit{\sqrthz}{\ensuremath{\sqrt{\text{\hertz}}}} 

\DeclareSIUnit\torr{torr}
\DeclareSIUnit\sccm{sccm}
\DeclareSIUnit\bar{bar}
\newcommand{\seeAppendix}[2]{Appendix~\hyperref[#2]{#1}}

\newcommand{\affilIFP}{Laboratory for Solid State Physics, ETH Zurich, CH-8093 Zurich, Switzerland.}
\newcommand{\affilMarc}{present address: attocube systems AG, 85540 Haar, Germany}

\begin{document}

\preprint{}

\title{Nanoscale magnets embedded in a microstrip}

\author{Raphael Pachlatko}\affiliation{\affilIFP}
\author{Nils Prumbaum}\affiliation{\affilIFP}
\author{Marc-Dominik Krass}\affiliation{\affilIFP}\affiliation{\affilMarc}
\author{Urs Grob}\affiliation{\affilIFP}
\author{Christian L. Degen}\affiliation{\affilIFP}
\author{Alexander Eichler}\affiliation{\affilIFP}

\date{\today}% It is always \today, today, but any date may be explicitly specified

\begin{abstract}
Nanoscale magnetic resonance imaging (NanoMRI) is an active area of applied research with potential use in structural biology and quantum engineering. The success of this technological vision hinges on improving the instrument's sensitivity and functionality. A particular challenge is the optimization of the magnetic field gradient required for spatial encoding, and of the radio-frequency field used for spin control, in analogy to the components used in clinical MRI. In this work, we present the fabrication and characterization of a magnet-in-microstrip device that yields a compact form factor for both elements. We find that our design leads to a number of advantages, among them a fourfold increase of the magnetic field gradient compared to those achieved with traditional fabrication methods. Our results can be useful for boosting the efficiency of a variety of different experimental arrangements and detection principles in the field of NanoMRI. 
\end{abstract}

%\pacs{}

\maketitle

%NOTE: spatial resolution is not up to standard with 10s integration time, these measurements are decoupled from each other!
%is there a reason we don't mention MRFM as a term at all?

% %%%%%%%%%%%%%%%%%%%%%%%%%%%%%%%%%%%%%%%%%%%%%%%%%%%%%%%%%%%%%%%%%%%%%%%%
%\section{Introduction}
% %%%%%%%%%%%%%%%%%%%%%%%%%%%%%%%%%%%%%%%%%%%%%%%%%%%%%%%%%%%%%%%%%%%%%%%%

A growing community of scientists is working towards nanoscale magnetic resonance imaging (NanoMRI) utilizing nanomechanical sensors~\cite{sidles1991,rugar_2004single,Degen_2009,poggio2010force,vinante2011magnetic,Nichol_2013,haas2022nuclear}, nitrogen-vacancy centers in diamond~\cite{Ajoy_2015,kost2015resolving,shi2015single,lovchinsky2016nuclear}, spin-dependent tunnelling~\cite{baumann2015electron,willke2018hyperfine,seifert2020,chen2022harnessing}, or superconducting resonators~\cite{Eichler_2017,bienfait2016reaching,Probst_2017}. The vision of NanoMRI is to enable three-dimensional, nondestructive imaging of individual molecules or solid-state samples with atomic or near-atomic spatial resolution~\cite{poggio2007nuclear, Degen_2009, grob_magnetic_2019}. Once this milestone is reached, many useful applications in structural biology or the design of quantum devices are within reach. However, detecting small ensembles of nuclear spins (e.g., hydrogen atoms) across distances of tens or hundreds of nanometers is an enormous challenge due to the small signal generated by each nuclear spin. The community is therefore actively seeking to optimize their measurement approaches.

\begin{figure*}[htb]
\includegraphics[width=0.75\textwidth]{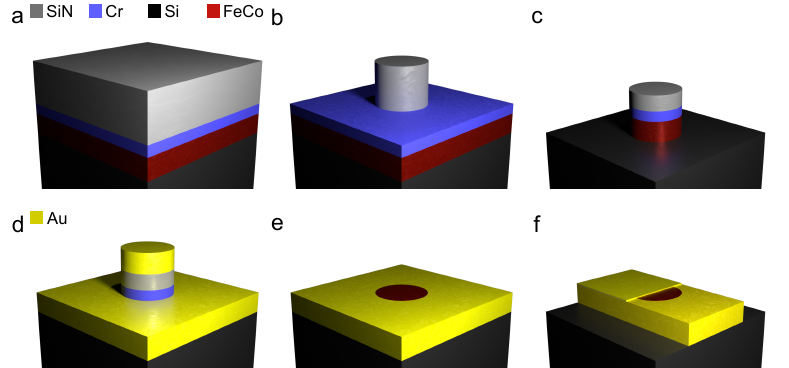}
\caption{Fabrication process for embedded nanomagnets: (a)~FeCo alloy (\SI{260}{\nm}) is evaporated on a Si wafer, followed by a spacer layer of Cr (\SI{130}{\nm}) and a SiN hardmask (\SI{550}{\nm}). (b)~The SiN film is etched into nanopillars using negative e-beam resist and reactive ion etching (RIE). (c)~Using the SiN nanopillars as hardmasks, the Cr and FeCo films are milled away in a variable-angle ion milling step. (d)~A Au film is evaporated and the surface is milled at a glancing angle. (e)~The Cr spacer is removed using Cr etch, followed by iterative polishing steps using glancing angle ion milling and additional Au layer evaporations. (f)~The SiN hardmask steps (a) through (c) are repeated to shape the microstrip from the Au film. The final \SI{15}{\nm} thick topping layer of Ti/Au is partially shown for visibility purposes.}
\label{fig:fig1}
\end{figure*}

\begin{figure}[htb]
\includegraphics[width=\columnwidth]{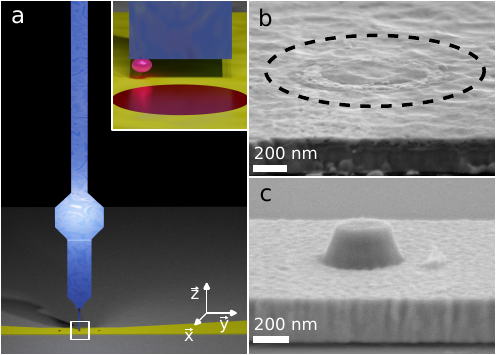}% size for 2column
\caption{Experimental arrangement. (a)~To-scale illustration of the cantilever (blue) and microstrip (yellow) assembly. An influenza virus (pink) is fixated to a Si nanorod at the cantilever end and positioned roughly \SI{100}{\nm} above the nanomagnet (red). The inset shows a zoom of the nanorod tip region marked by a white square. (b)~Scanning electron micrograph of magnet-in-microstrip device. The dotted circle denotes its position within the gold surface. (c)~Scanning electron micrograph of a traditional magnet-on-microstrip device, consisting of a FeCo nanomagnet in a truncated cone shape on top of a Au microstrip.%The nanomagnets have diameters of \SI{687}{\nm}, \SI{671}{\nm} and\SI{503}{\nm} and are spaced apart \SI{5}{\um}.
}
\label{fig:fig2}
\end{figure}

NanoMRI relies on the simultaneous application of strong magnetic field gradients for spatial resolution and radio-frequency fields for coherent spin control. Field gradients can be generated by means of nanoscale electromagnets or permanent magnets~\cite{rugar_2004single,Nichol_2012,longenecker2012high,mamin2012,grob_magnetic_2019}, while current-carrying elements like microstrips are used to produce RF fields for spin manipulation ~\cite{poggio2007nuclear,sasaki2016broadband}. For optimal signal generation, both components (field gradient source and microstrip) should be placed as close to the sample as possible. This geometrical optimization can be hard to achieve in a scanning probe setup, especially when the instrument design is constrained by the necessity of working in vacuum, at low temperatures, or in large external magnetic fields.

In this paper, we demonstrate on-chip integration of a nanoscale field gradient source and a microstrip resulting in a compact geometry. By embedding the nanomagnet within a gold (Au) microstrip, we can place an influenza virus test sample within tens of nanometers from both elements, enabling optimal static and RF field strengths. We test the performance of our magnet-in-microstrip device in a force-detected NanoMRI setup and find that the field gradient in $x$-direction, the direction of the cantilever oscillation, is boosted by a factor of $\sim$4 relative to magnet-on-microstrip devices~\cite{grob_magnetic_2019}. As a consequence, we detect nuclear spin ensembles within an averaging time of ten seconds, roughly one order of magnitude faster than in previous work~\cite{Krass_2022}. We further demonstrate a best-effort spatial resolution below \SI{1}{\nano\meter} in the direction of oscillation. Our geometry offers a general pathway to improving the signal acquisition in force-detected NanoMRI.

% %%%%%%%%%%%%%%%%%%%%%%%%%%%%%%%%%%%%%%%%%%%%%%%%%%%%%%%%%%%%%%%%%%%%%%%%
%\section{Setup and Fabrication}
% %%%%%%%%%%%%%%%%%%%%%%%%%%%%%%%%%%%%%%%%%%%%%%%%%%%%%%%%%%%%%%%%%%%%%%%%

The process to create a microstrip with embedded nanomagnets is summarized in Fig.~\ref{fig:fig1} and explained in detail in the the appendix. In short, the fabrication follows a top-down approach with a silicon nitride (SiN) hardmask to pattern a FeCo layer (FeCo, 65$\%$/35$\%$ at.wt.) into the desired cylinder shape. In a second step, a Au layer is applied, the hardmask is removed, and the nanomagnet and Au film are ion milled at a glancing angle to create a planar surface (typical root-mean-square variations on the order of \SI{7}{\nm})~\cite{Yoo2018a}. In a third step, the Au layer is patterned into a microstrip using a second hardmask. Finally, the microstrip and nanomagnet are covered with an additional thin Ti/Au layer (\SI{15}{\nm}) to minimize electrostatic work function differences across the surface.

To test the functionality of the embedded nanomagnets, we mount them in a magnetic resonance force microscopy (MRFM) setup~\cite{Degen_2009,poggio2010force,Krass_2022}. Our setup is suspended in a vacuum chamber inside a liquid helium bath cryostat, with a base temperature of $T = \SI{4.7}{\kelvin}$ and a pressure of $p = \SI{1e-7}{\milli\bar}$. The force sensor is an in-house fabricated silicon cantilever, see Fig.~\ref{fig:fig2}(a)~\cite{mamin2001,moores_2015accelerated,grob_magnetic_2019,Krass_2022}.
The cantilever has dimensions \SI{150}{\micro\meter} $\times$ \SI{4}{\micro\meter} $\times$ \SI{130}{\nano\meter}, a free resonance frequency $f_0 = \SI{3500}{\hertz}$, a mass $m = \SI{100}{\pico\gram}$, and a quality factor $Q = 25000$ at cryogenic temperatures around \SI{4}{\kelvin}. These properties result in an intrinsic spring constant of $k_0 = \SI{50}{\micro\newton\per\meter}$ and a damping coefficient of $\gamma_0 = \SI{8.8e-14}{\kilo\gram\per\second}$. Sensor readout is achieved using a fiber-optic interferometer~\cite{rugar1989,heritier_nanoladder_2018}. By analyzing the AC and DC components of the reflected light, we extract the cantilever frequency and static deflection, respectively. The laser temperature is stabilized to within \SI{10}{\milli\kelvin} to reduce signal drift~\cite{Krass_2022}. Samples are attached to the end of the cantilever via a silicon nanorod~\cite{Krass_2022,overweg2015probing} and then brought into close proximity with the nanomagnet on the chip. A SEM micrograph of the magnet-in-microstrip device, which forms the main advance of this work, is shown in Fig.~\ref{fig:fig2}(b) together with a traditional magnet-on-microstrip device, shown in Fig.~\ref{fig:fig2}(c).

%To attach the influenza virus sample to the cantilever, silicon nanorods with a bottom cross-section of $\SI{500}{\nm}\times\SI{500}{\nm}$ are fabricated on waferscale in a cleanroom to serve as fixation substrate for the virus sample. Individual chips with multiple nanorods present are immersed in a buffered virus solution with a suitable concentration to result in isolated deposited viruses on the nanorod surface. Subsequently, the viruses are fixated using a glutaraldehyde/formaldehyde process and stained using osmium tetroxide. The nanorod is then manually glued to the cantilever tip using a set of micromanipulators and epoxy glue.

%The nanomagnet, made from iron-cobalt alloy (FeCo, 65$\%$/35$\%$ at.wt.),
%and possessing a diameter of roughly \SI{670}{\nm} produces a large magnetic field gradient. The microstrip is made from Au and serves as the carrier for RF current pulses~\cite{Degen_2009}. Traditionally, the nanomagnet is fabricated on top of the microstrip in a two-step evaporation and lift-off process, cf. Fig.~\ref{fig:fig1}(b)~\cite{poggio2007nuclear}. While this process is comparably straightforward, it results in conical nanomagnets with rounded edges, which limits the achievable field gradients~\cite{grob_magnetic_2019}. In addition, we found that the cantilever strongly bends towards the nanomagnet edges due to non-contact interactions~\cite{Krass_2022}.

\begin{figure*}[htb]
\includegraphics[scale = 1]{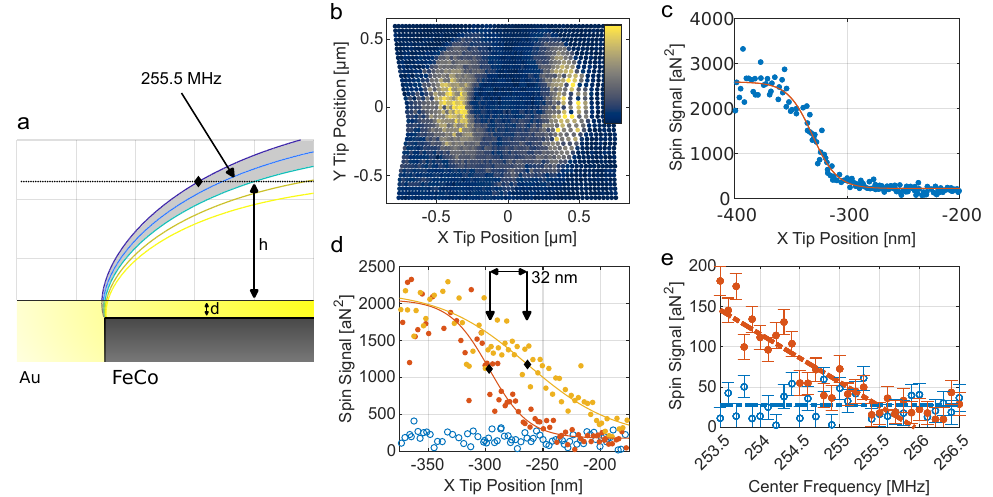}
\caption{Device characterization using MRFM. (a)~Simulated spatial distribution of magnetic isolines in a vertical cut through the magnet center. Colored lines indicate positions with constant Larmor frequency $f_\mathrm{L}$, separated by \SI{1}{\mega\hertz} each. The shaded area corresponds to a resonant slice between $f_\mathrm{L} = \SI{256.5}{\mega\hertz}$ and \SI{254.5}{\mega\hertz}. The grey raster lines are separated by \SI{50}{\nano\meter}. The solid black line indicates the standoff height $h$ at which the magnetization of the nanomagnet was determined in (e). The tip-magnet distance is equal to $h+d$ with $d=\SI{15}{\nm}$. (b)~Two-dimensional MRFM scan at $h = \SI{100}{\nm}$ above the surface with an integration time of \SI{60}{\second}, a center frequency of \SI{245}{\mega\hertz}, and a frequency deviation of \SI{800}{\kilo\hertz} (half peak-to-peak frequency chirp range). %The positions of the data points are plotted at their true position, taking into account static deflections of the cantilever~\cite{Krass_2022}.
The colorscale ranges from 0 to \SI{1500}{\atto\newton^2} with higher values capped to improve the visibility of the signal regions. (c)~One-dimensional MRFM scan measured at $h = \SI{50}{\nm}$ with identical parameters to (b). The signal rise is fitted to a hyperbolic tangent step function. (d)~One-dimensional MRFM scans measured at $h = \SI{20}{\nm}$ with \SI{60}{\second} integration time and \SI{1}{\mega\hertz} pulse bandwidth. The center frequencies shown are \SI{245}{\mega\hertz} at zero inversion bandwidth (open blue points), \SI{245}{\mega\hertz} at \SI{1}{\mega\hertz} inversion bandwidth (red), and \SI{248}{\mega\hertz} at \SI{1}{\mega\hertz} inversion bandwidth (yellow). Black squares mark the positions where the signal reaches \SI{50}{\percent} of its maximal value for each center frequency. %Using the physical spacing between these points (\SI{32}{\nm}), we can determine the magnetic field gradient in this specific location to be \SI{2.1}{\mega\tesla\per\meter}.
(e)~Static MRFM spectrum measured at $h = \SI{100}{\nm}$, shown as a black line in (a). The signal integration time was \SI{600}{\second} with a standard inversion bandwidth of \SI{1}{\mega\hertz}. %The intersection of a conservative linear fit of the signal decay (red) with the background noise (blue) yields a cutoff frequency at \SI{255.5}{\mega\hertz} and a conservative estimation of the magnetization around \SI{2.0}{\tesla}.
}
\label{fig:fig3}
\end{figure*}

%upper estimate on spectrum is 259.4 MHz, leading to an upper estimate on magnetization of 1.85T (not sure if we need the exact contour line for this)
\begin{figure*}[htb]
\includegraphics[scale = 1]{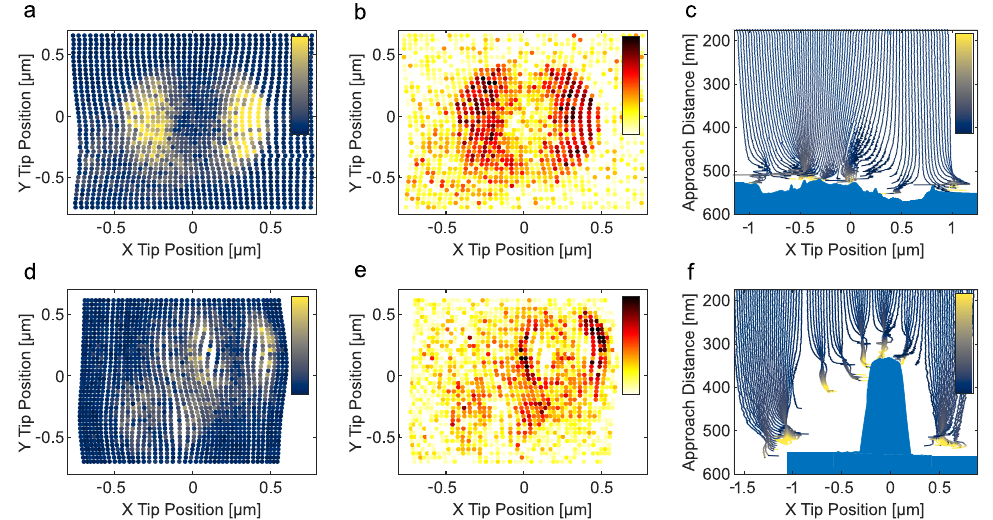}% size for 2column
\caption{Direct comparison of magnet-in-microstrip (a-c) and magnet-on-microstrip (d-f) devices. Both scans image an identical virus sample~\cite{Krass_2022}. (a)~Two-dimensional MRFM measurement recorded with the magnet-in-microstrip device with an integration time of \SI{10}{\second} and a $h= \SI{100}{\nm}$. The colorscale ranges from 0 to \SI{1000}{\atto\newton^2} with higher values capped to improve the visibility of the signal regions. An embedded nanomagnet %located within \SI{5}{\micro\meter} (i.e., located on the same microstrip
with a diameter of ca. \SI{500}{\nm} was used for this dataset.
(b)~Signal-to-noise ratio (SNR) for the data shown in (a). The colorscale ranges from 1 (bright) to 8 (dark) with higher SNR values capped to illustrate signal shells characteristic for this nanomagnet shape. 
(c)~Approach trajectories of the cantilever towards the magnet-in-microstrip sample. The blue shape indicates the surface topography as determined in a separate atomic force microscopy scan. Approximately vertical lines correspond to individual trajectories starting from the top and terminating when either the frequency shift of amplitude drop triggers a stop condition. The color scale of the trajectories indicates the measured frequency, ranging from \SI{3}{\kilo\hertz} (dark blue) to \SI{9}{\kilo\hertz} (yellow). 
(d)~Two-dimensional MRFM measurement recorded with a magnet-on-microstrip device~\cite{Krass_2022} with a signal integration time of \SI{60}{\second} and $h= \SI{100}{\nm}$. The colorscale ranges from 0 to \SI{500}{\atto\newton^2} with higher values capped to improve the visibility of the signal regions. The cantilever and influenza sample are identical to the ones used in this work. (e)~SNR for the data shown in (d). The colorscale is identical to the one in (b).
(f)~Approach trajectories of the cantilever towards the magnet-on-microstrip sample, analogous to (c).}
\label{fig:fig4}
\end{figure*}
% \begin{figure}[htb]
% \includegraphics[width=\columnwidth]{figures/Fig5.pdf}% size for 2column
% \caption{}
% \label{fig:fig5}
% \end{figure}

% %%%%%%%%%%%%%%%%%%%%%%%%%%%%%%%%%%%%%%%%%%%%%%%%%%%%%%%%%%%%%%%%%%%%%%%%
%\section{Results}
% %%%%%%%%%%%%%%%%%%%%%%%%%%%%%%%%%%%%%%%%%%%%%%%%%%%%%%%%%%%%%%%%%%%%%%%%
% gradient data: 3MHz spacing at 42.577 MHz/T for protons, physical slice spacing between slope maxima = 32 nm yields 2.2019 MT/m! -> no longer accurate, .txt files take single data values, need averaged values to yield Gx of 2.1 MT/m; small change, but important

In MRFM, the magnetic field gradient $G_x = \partial B_z/\partial x$ in the direction of the mechanical sensor displacement $x$ serves two purposes: (i) it enables spatial selectivity, and (ii) it transduces a spin's $z$-magnetization $\mu_z$ into a force $F_x=\mu_z G_x$ that is measured with the mechanical sensor (i.e., the cantilever). Coherent inversion of the spins with a rate equal to $2 f_0$ leads to a force at $f_0$, thereby driving the cantilever into resonant oscillation and maximizing the readout efficiency. 
We use rapid adiabatic passage pulses with hyperbolic secant modulation for high-fidelity spin inversions~\cite{grob_magnetic_2019,tannus1996}. The pulse bandwidth is tuned to invert spins within a selected narrow range of Larmor frequencies centered around a carrier frequency referred to as ``center frequency'', which translates into a ``resonant slice'' in real space, cf. Fig.~\ref{fig:fig3}(a). The sharpness of the pulse window function, i.e., the distance over which the pulse inversion efficiency drops from ``on'' to ``off'', determines the spatial resolution~\cite{Degen_2009,grob_magnetic_2019}. A static magnetic field $B_0 = \SI{5.6}{\tesla}$ applied in $z$ direction serves to magnetize the nanomagnet and to provide a quantization axis for the spins.

In our first experiment, we scan the influenza sample over the embedded nanomagnet at a constant height and invert the proton spins ($^1$H nuclei) contained therein in a Larmor frequency range of $\SI{244}{\mega\hertz}$ to $\SI{246}{\mega\hertz}$, cf. Fig.~\ref{fig:fig3}(a). Measuring the force generated by the inverted spins in the sample, we obtain the double crescent shape characteristic for this protocol~\cite{Degen_2009}, cf. Fig.~\ref{fig:fig3}(b). The dependence of the signal strength on the scan position defines a one-dimensional spatial resolution~\cite{grob_magnetic_2019}. In Fig.~\ref{fig:fig3}(c), the maximum slope of the hyperbolic tangent function is compared to the standard variation $\sigma = \SI{47.3}{\atto\newton\squared}$, yielding a resolution of \SI{0.91}{\nm}. This value is on par with the value reported previously with a magnet-on-microstrip device in a carefully optimized setting~\cite{grob_magnetic_2019}.

In a next step, we characterize $G_x$ by testing that the signal from different resonant slices corresponds to different spatial regions. In Fig.~\ref{fig:fig3}(d), we plot the results of two lateral scans with center frequencies differing by $\Delta{f} = \SI{3}{\mega\hertz}$. We find that the two slices are shifted by a distance of $\Delta{x} = \SI{32}{\nm}$, corresponding to $G_x = \frac{\Delta{f}}{\gamma \Delta{x}} = \SI{2.1}{\mega\tesla\per\meter}$, where $\gamma = \SI{42.58}{\mega\hertz\per\tesla}$ is the gyromagnetic ratio of protons. This is roughly four times larger than the best value ($G_x = \SI{0.56}{\mega\tesla\per\meter}$) we measured with the traditional magnet-on-microstrip design in our setup, which are evaporated through resist masks~\cite{grob_magnetic_2019}. (Note that the gradient can be further increased by using other materials~\cite{longenecker2012high,mamin2012}.) We ascribe the improvement in $G_x$ to the sharper magnet edge and steeper magnet sidewalls achieved through the ion-milling-based fabrication process. %The patterning used for the embedded nanomagnets has an edge sharpness limited mainly by the mask hardness, while traditional devices feature magnets evaporated through resist masks. Evaporation produces naturally rounded edges as the mask slowly narrows during the material deposition.

%However, the gradient that is directly relevant for the signal strength in our experiment is $G_x$, for which we are not aware of a higher value reported in literature. %We note that the measured $G_x$ values are significantly lower than simulated ones, which we attribute to the underestimation of edge rounding and layer passivation~\cite{longenecker2012high}.

The magnetization of the nanomagnet is estimated from the signal onset at the high-frequency edge of the slice~\cite{mamin2007}. First, we perform scans at a fixed height above the nanomagnet.
%, see Supporting Information (SI) ... for details. 
This scan allows us to determine the center of the nanomagnet. The magnetization is then determined by positioning the cantilever in a signal region along the x-axis with the distance from the nanomagnet center known, and acquiring an MRFM spectrum~\cite{longenecker2012high}, shown in ~\ref{fig:fig3}(e). From the center frequency-dependent signal decay, we calculate a conservative saturation magnetization of \SI{2.0}{\tesla}.%, see SI ... for details. 
This magnetization is about \SI{9}{\percent} higher compared to other FeCo nanomagnets measured in the same setup~\cite{Krass_2022}, indicating excellent material properties.

% %%%%%%%%%%%%%%%%%%%%%%%%%%%%%%%%%%%%%%%%%%%%%%%%%%%%%%%%%%%%%%%%%%%%%%%%
%\section{Discussion}
% %%%%%%%%%%%%%%%%%%%%%%%%%%%%%%%%%%%%%%%%%%%%%%%%%%%%%%%%%%%%%%%%%%%%%%%%

Our measurements indicate several immediate advantages of the magnet-in-microstrip over the magnet-on-microstrip design. Firstly, the much larger $G_x$ directly improves the speed and signal-to-noise ratio (SNR) of NanoMRI scans. In MRFM, the SNR is proportional to $G_x^2$~\cite{Degen_2007_role}. An improvement by a factor of four in $G_x$ therefore translates into an SNR improvement by a factor of 16 per spin under ideal conditions. However, an increase in $G_x$ leads to physically thinner resonance slices, reducing the addressed spin sample volume and therefore also the signal. As a consequence, we observe an SNR value that is only 2 to 3 times greater than previously measured with magnet-on-microstrip devices.
%reported (see SNR line scan and adiabaticity plot in the SI). 
 
%The improvement in the measured $G_x$ is important for a variety of reasons. The bare FeCo stems from the same source and is evaporated in the same machine, resulting in the same magnetization within the measurement uncertainty.

%to the sharper magnet edge obtained through the different fabrication process as well as steeper magnet sidewalls leading to a sharper angle at the nanomagnet edge. 

%Numerical finite-element simulations show that a decrease in edge rounding from \SI{40}{\nm} to \SI{5}{\nm} increases $G_x$ measured at a distance of \SI{20}{\nm} by \SI{23}{\percent} (see SI).

We can leverage the high SNR to reduce the acquisition time per data point.
%for high detection bandwidth, i.e., shorter acquisition times. 
In Fig.~\ref{fig:fig4}(a) and (b), we show a two-dimensional spatial scan measured with an integration time of only \SI{10}{\second} per pixel, compared to typical integration times between \SI{60}{\second} and \SI{300}{\second} in earlier measurements~\cite{moores_2015accelerated,grob_magnetic_2019,Krass_2022}. The SNR we obtain reaches values above 8 over a wide range of the scan. The signal strength and SNR clearly surpass the results obtained for the same sample with a magnet-on-microstrip device~\cite{Krass_2022}, where a sixfold longer integration time and smaller tip-magnet distance were used, see Fig.~\ref{fig:fig4}(d) and (e). This speedup, enabled by the large $G_x$ values, will be valuable when performing large three-dimensional scans of extended nanoscale samples.

A further advantage of the magnet-in-microstrip design is the reduction of topographical features which drastically reduces static cantilever bending close to the surface. In Fig.~\ref{fig:fig4}(c) and (f), we compare the surface approach trajectories measured over the two device types. In contrast to the traditional nanomagnets, whose profile is found to induce significant bending and feedback instabilities~\cite{Krass_2022}, the nearly seamless surface of the embedded magnets allows for faithful scanning down to a few tens of nanometers surface promimity and facilitates data acquisition at crucial positions close to the magnet edges.

%The embedded magnet design also reduces the separation between NanoMRI samples and the microstrip surface, which potentially allows working with smaller pulse currents. For instance, we show in the SI that higher spin inversion fidelity is reached for the same current with embedded magnets than with magnet-on-microstrip devices. This feature will become significant when performing NanoMRI measurements at millikelvin temperatures where the heating due to the current is a limiting factor.

Finally, a recent study highlighted the connection between topographical features and non-contact friction (NCF)~\cite{Heritier_2021} via dielectric interactions~\cite{kuehn_dielectric_2006, yazdanian_dielectric_2008}. We expected that the embedded magnet design could be advantageous for obtaining lower NCF. However, contrary to our expectations, we generally observed an increase of NCF compared to magnet-on-microstrip devices. We also found a strong dependence of the NCF on the top layer material, with devices capped by a Au layer performing significantly better than others where the magnet is left exposed. %This will be a crucial point to be addressed in future studies.
Looking forward, we speculate that the NCF could be further reduced by covering the magnet and the Au microstrip with a conducting, atomically flat layer. Such two-dimensional materials are being researched intensively for applications in wearable electronics~\cite{orts2022electrically} and offer many advantages in various areas of nanotechnology. Besides screening unwanted electrical interactions, a layer consisting of only a few atomic layers would also allow a closer approach of the scanning tip to the magnet surface and make larger gradients accessible. Under such conditions, we expect that the magnet-in-microstrip design would result in an up to tenfold improvement of $G_x$. Applied to MRFM, a tenfold higher $G_x$ would translate into a hundredfold increase in signal strength and SNR or a reduction in averaging time by four orders of magnitude at constant SNR. The magnet-in-microstrip design can therefore significantly boost the performance of NanoMRI instruments.

%DANKSAGUNG!!
\section*{Acknowledgments}
We thank the operations team of the FIRST cleanroom, especially Sandro Loosli and Petra Burkard, as well as the operations team of the Binnig and Rohrer Nanotechnology Center (BRNC) at IBM Rüschlikon, especially Ute Drechsler, Richard Stutz and Dr. Diana Dávila Pineda, their continued support and expertise has enabled the development of this new device generation.

\appendix
\section{Appendix A: Fabrication of Embedded Nanomagnets}\label{appendix:FabAppendix}

%This appendix section serves to detail the fabrication process to a degree where prototyping and eventual replication can be achieved with equivalent equipment. It should be noted however that many processes involving reactive Ion etch (RIE), Ion Beam Etching (IBE), also referred to as Ion Milling,Plasma Enhanced Chemical Vapor Deposition (PECVD) and virtually any step in the fabrication is influenced by specific device parameters and must be calibrated in-situ to any new devices. This particularly applies to etch rates in RIE systems as well as milling rates at various angles and powers for IBE. Chemicals and resists will be listed without greater detail under the assumption that a future user can familiarize themselves.

\subsection{Wafer Preparation}

The fabrication process begins at the wafer scale in order to streamline the initial processes. The wafers used were \SI{10}{\cm}  (4-inch) silicon wafers made by MicroChemicals GmbH (article number WTD40525250B1314S102, boron-doped, \SIrange{1}{10}{\ohm\cm}, \SI{100}{\nm} oxide layer). In order to ensure consistent alignment between multiple fabrication steps, markers are etched into the wafer itself to provide physical landmarks. While these markers can be seen optically, their main purpose is to enable visibility under SEM.

\begin{itemize}
    \item Using Laser Writing (DWL 66+, Heidelberg Instruments) and the optical resist AZ 1518 (MicroChemicals GmbH), a periodic grid of alignment markers and chip borders is exposed into the resist layer on top of the wafer and developed using AZ 400K 1:4 MIC developer (MicroChemicals GmbH). Standard hardbake and cleaning measures are performed.
    \item Using a two-stage RIE etch process (PlasmaPro 80 RIE, Oxford Systems), the silicon oxide layer is breached and subsequently the underlying silicon is etched to a depth of roughly \SI{2}{\um}. The process step for the oxide breach uses CF$_4$ at \SI{150}{\watt}, \SI{65}{\milli\torr}, \SI{20}{\sccm}  for \SI{2.5}{\minute}. The process step for the silicon etch uses SF$_6$ \& O$_2$ at \SI{100}{\watt}, \SI{100}{\milli\torr}, \SI{50}{\sccm}/\SI{10}{\sccm} for \SI{2}{\minute}. %Power and flow rates will vary for different systems and will need to be configured accordingly.
    Etch depth into the silicon is not relevant as long as it is within micron range and does not deform the markers in the process. After the RIE etch, the resist is lifted and the wafer is cleaned in solvent baths of acetone and isopropyl alcohol as well as in a mild oxygen plasma.
    \item Using a BAK 501 LL electron beam evaporator (Evatec AG), \SI{10}{\nm} of titanium and \SI{260}{\nm} of FeCo alloy (65$\%$/35$\%$ at.wt.) are evaporated onto the wafer surface without any masking to achieve a uniform film. The titanium serves as adhesion layer, while the FeCo will form the nanomagnets. Once the evaporation is complete, the wafer is subjected to glancing angle IBE in order to further planarize the surface and remove surface impurities, removing ca. \SI{10}{\nm} of material in the process. Once complete, \SI{3}{\nm} of platinum are evaporated onto the FeCo layer as a protection against oxidation, followed by \SI{130}{\nm} of Cr, which serves as a spacer layer for the following process.

\end{itemize}

\subsection{Hardmask Fabrication}

To shape the nanomagnets from the bulk FeCo film, Argon Ion Beam Etching is used, with Cr acting as a spacer layer and a hardmask made from silicon nitride (SiN) acting as physical shield from the argon ion impacts. Before the hardmask is grown, the wafer is diced into smaller chips ($10\times \SI{9}{\mm}$) for ease of handling and prototyping.

\begin{itemize}
    \item Using PECVD, a SiN layer with a thickness of \SI{550}{\nm} is grown on top of the Cr layer. The step is performed at a process temperateure of \SI{300}{\celsius}.
    \item In order to pattern the SiN layer, a negative resist (ma-n 2410, micro resist technology GmbH) is applied and exposed via electron beam lithography (EBL). Prior to resist application, the chips need to be washed thoroughly in acetone and isopropyl alcohol to improve resist adhesion to the SiN film.
    \item Using the engraved alignment markers, the core positions on the chip are located and exposed with simple circular patterns to form resist pillars. Three different magnets are implemented on each microstrip with a spacing of \SI{5}{\um} between the centers. The radii written in the exposure pattern are \SI{250}{\nm}, \SI{250}{\nm}, and \SI{325}{\nm}, with corresponding exposure doses of \SI{350}{\micro\coulomb\per\cm^2}, \SI{787.5}{\micro\coulomb\per\cm^2}, and \SI{350}{\micro\coulomb\per\cm^2}. The doses and radii control the width of the resulting resist pillars, and therefore the sizes of the nanomagnets. Development of the resist is achieved with MF 319, a TMAH-based developer.
    \item Using a two-step RIE process, the resist pattern is then transferred to the SiN film, resulting in the desired SiN hardmask shape. The RIE steps consist of a \SI{180}{\second} etch with \SI{30}{\sccm} of SF$_6$ at \SI{150}{\watt} and \SI{30}{\milli\torr}, followed by a \SI{225}{\second} etch with \SI{50}{\sccm}/\SI{5}{\sccm} of CHF$_3$/O$_2$ at \SI{100}{\watt} and \SI{55}{\milli\torr} (the second step employs a mixed process gas). The first etch step produces sloped sidewalls while the second produces nearly vertical sidewalls. A combination of the two different etch steps assures an optimized etch profile for IBE.
    \item Once the RIE process is complete, the negative resist is lifted in a solvent bath of NMP or DMSO at \SI{80}{\celsius}. NMP is more effective at lifting quickly and cleanly, however DMSO, which is nontoxic, and has also yielded satifactory results.
    
\end{itemize}

\subsection{IBE-assisted Creation of Nanomagnets}

Now that the hardmask is in place, Ion Milling of the metallic substrate can begin.

\begin{itemize}
    \item Using an Ionfab 300 Plus (Oxford Instruments), the entire metallic film surrounding the SiN hardmask is eroded with argon ion bombardment at a beam current of \SI{500}{\milli\ampere} while the sample holder is rotated around its central axis. The critical issue in this process is to achieve both an optimal pattern transfer of the hardmask while simultaneously mitigating redeposition. To that end, the incidence angle of the ion beam is varied during the milling step, and the material ejection is monitored using secondary ion mass spectroscopy (SIMS). The process begins with an incidence angle of \SI{20}{\degree} (from the surface normal, \SI{0}{\degree} being perpendicular to the surface) for \SI{21}{\minute}, eroding \SI{130}{\nm} of Cr, \SI{3}{\nm} of Pt and most of the \SI{250}{\nm}, approximately \SI{235}{\nm}.
    \item In order to remove redeposited material from the newly created metal pillars' sidewalls, the incidence angle is changed to \SI{80}{\degree} for \SI{10}{\minute} at constant rotation. SIMS readout at such glancing angles is impaired and milling rates for such angles are best calibrated using test- and sacrificial samples.
    \item The steep \SI{20}{\degree} angle is resumed for the final step until the elemental signal of the Ti adhesion layer diminishes in the SIMS readout, which typically occurs after around \SI{100}{\second}. This decrease marks the breach of the final metal layer and the completion of the nanomagnet stacks.
    \item The shape of the nanomagnets is examined using AFM and SEM; to that end, a sacrificial sample is immersed in Cr etch (Cerium Nitrate solution) in order to selectively dissolve the Cr spacer separating the hardmask and the newly created nanomagnet. The existence of the spacer allows the selective separation of the nanomagnet from its hardmask, however the mask needs to remain at this stage for regular samples. Using SEM, the milling outcome and general shape of the nanomagnet is examined, while AFM yields the precise height of the nanomagnet with respect to the silicon surface. In the specific example of the sample measured in this publication, the height of the nanomagnets was determined to be \SI{260}{\nm}.
\end{itemize}

\subsection{Au Layer Evaporation}

Using the determined height of the nanomagnet, a variable-angle evaporation of Ti and Au is conducted to evaporate the Au film onto the silicon surface, thereby covering the free-standing nanomagnet arrays with a Ti/Au metal film. In order to monitor the process, 4 sacrificial chips are created before the evaporation and covered with chromium as a stopping layer. These chips are then coated along with the regular samples.

The coating process needs to leave the chromium spacer exposed for a future liftoff. The sides of the nanomagnets need to be coated homogeneously, ideally in a relatively passive material, and the final Au film needs to match the height of the nanomagnet as closely as possible to minimize topographic features.

\begin{itemize}
    \item Starting the evaporation at an angle of \SI{60}{\degree} to the surface normal with a rotating sample holder, \SI{20}{\nm} of Ti is evaporated onto the surface of the chip as well as onto the sidewalls of the nanomagnet arrays. The effective layer thickness on the chip surface due to tilt is \SI{10}{\nm}. The Ti film serves as a crucial adhesion film for the sidewalls of the nanomagnets as well, ensuring no gaps will form between the FeCo and Au films. This is vital to avoid galvanic corrosion effects which are facilitated by trapped moisture in potential gaps, an effect which has been observed on previous device iterations.
    \item Using the same incidence angle, \SI{100}{\nm} of Au is evaporated, homogeneously coating the surface and sidewalls in a (in case of the sidewalls relatively thin) layer of Au. The effective Au layer thickness on the chip surface due to tilt is \SI{50}{\nm} at this point.
    \item With an incidence angle of \SI{0}{\degree}, \SI{100}{\nm} Au are evaporated onto the chip, translating directly to effective layer thickness and mostly leaving the sidewalls unaffected.
    \item In a third evaporation step, with an incidence angle of \SI{30}{\degree}, \SI{173}{\nm} of Au is evaporated. This angle is chosen to minimize shadowing effects from the rotating evaporation while still maintaining coverage of the sidewalls.The effective layer thickness on the chip surface due to tilt is \SI{150}{\nm}.
    \item Using a sacrificial chip introduced previously, the Au layer is masked with AZ 1518 analogously to the marker fabrication, and is etched down with RIE using CHF$_3$/O$_2$ with a flow of \SI{45}{\sccm}/\SI{2}{\sccm} for \SI{13.5}{\minute} at \SI{300}{\watt} and \SI{25}{\milli\torr}. Conclusion of the etch is verified in SEM and the layer thickness is confirmed to be \SI{300}{\nm} via AFM, yielding a total thickness of \SI{310}{\nm} when including the Ti adhesion layer.
\end{itemize}

\subsection{Hardmask Liftoff and Polishing}

Once the height of the Au film has been determined, the distance between the Au layer surface and the nanomagnet surface can be calculated to be \SI{50}{\nm} in total. It is generally easier to overshoot the Au layer thickness and then erode the excess than the reverse case, as the following steps show:

\begin{itemize}
    \item Using the same CHF$_3$/O$_2$ RIE recipe with a shorter runtime of \SI{107}{\second}, \SI{40}{\nm} of Au are removed in an anisotropic vertical etch, leaving \SI{10}{\nm} of Au layer above the nanomagnet level. Note that vertical IBE has proven to be unreliable to remove material in a strictly vertical etch without shadowing, which in this case must be avoided. Furthermore, care must be taken not to overetch the layer, as a breach of the nanomagnet sidewalls would be catastrophic, corroding the material in the process.
    \item In order to remove the SiN hardmasks and expose the nanomagnets, the sidewalls of the Cr spacer on top need to be exposed. To do so without eroding the surrounding Au film, \SI{10}{\nm} Cr and \SI{20}{\nm} Ti are evaporated vertically as a protection layer and the whole chip is ion milled at \SI{80}{\degree} incidence angle for \SI{18}{\minute}, which roughly is the time the protection layer can withstand the ion milling at this angle.
    \item After the glancing angle ion milling step, the protection layer and the SiN hardmasks are removed via Cr etch and the topography is examined in SEM and AFM.
    \item In case of overshooting the desired milling depth, a generous layer of Au/Ti can be applied and polished down iteratively to avoid another overshoot.
    \item In case of insufficient milling depth, an application of another Cr/Ti protection layer and a further milling step can reduce the layer thickness further without compromising topography or the nanomagnet itself. Protection layers are especially desirable if the topography is indufficiently flat, otherwise a simple glancing angle milling step can be sufficient.
    \item Once the nanomagnets are visible below the Au layer in SEM, gentle polishing steps can be made to remove any remaining Au islands on top of the nanomagnet. This is made easier by the comparably high milling rate of Au compared to FeCo at all angles.
    \item For the embedded nanomagnets measured in this paper, 5 iterative polishing steps have produced a suitable result.
\end{itemize}

\begin{figure*}[t]
\includegraphics[scale = 1]{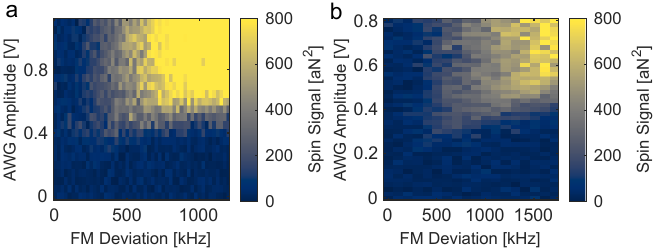}% size for 2column
\caption{Comparison of spin inversion pulse characteristics between microstrip device generations for \SI{60}{\second} integration time. Color scale bars are limited to \SI{800}{\atto\newton^2} for better low-signal contrast. (a) Spin signal recorded with the magnet-in-microstrip device from Fig.~2(b) at a distance of \SI{90}{\nm} above the microstrip surface (sample-magnet separation of \SI{110}{\nm}). High signal is obtained roughly above \SI{400}{\kilo\hertz} FM deviation and \SI{0.5}{\volt} AWG amplitude, with a measured maximum of \SI{2000}{\atto\newton^2}. (b) Spin signal recorded with a conventional magnet-on-microstrip design from Fig.~2(c) at a sample-magnet separation of \SI{30}{\nm}. The maximum measured spin signal is \SI{900}{\atto\newton^2}.
}\label{fig:AdiaAppendix}
\end{figure*}

\subsection{Microstrip Creation from Au film}

The current state of the sample chip is a planar Au film with FeCo nanomagnets embedded in its surface. In order to pattern a microstrip from the bulk film, the following steps are taken:

\begin{itemize}
    \item Using e-beam evaporation and PECVD, a \SI{130}{\nm} thick Cr film and a \SI{250}{\nm} thick SiN film are applied to the sample chip.
    \item Using EBL, a resist bilayer of 50K/950K PMMA (Allresist GmbH) is exposed with a dose of \SI{200}{\micro\coulomb\per\cm^2}. Alignment is again achieved using the etched alignment markers in the bulk substrate.
    \item The resist is developed using MIBK developer and \SI{40}{\nm} of Cr are evaporated onto the now exposed SiN layer, forming a Cr film in the shape of the desired microstrip, positioned above the nanomagnet array buried below. The resist is then lifted using a solvent bath.
    \item The Cr layer acts as etch ask for a RIE etch step to etch the SiN film using the CHF$_3$/O$_2$ recipe mentioned before, forming the hardmask for the next milling step. A hybrid etch recipe is not critical in this case.
    \item The chip is ion milled at \SI{20}{\degree} for a relatively short \SI{10}{\minute} until Ti is visible in the SIMS, signalling the removal of the Au layer, then polished at \SI{80}{\degree} for \SI{5}{\minute} to remove redeposited material from the microstrip sidewalls and the Cr spacer sidewalls. A final \SI{2}{\minute} are spent milling at \SI{20}{\degree} to breach the Ti layer and expose the surrounding silicon oxide substrate.
    This concludes the patterning of the microstrip and allows the SiN hardmask to be lifted using a Cr etch.
    \item A final layer of Ti/Au with thickness of \SI{3}{\nm} and \SI{12}{\nm}, respectively, is applied identically to the \SI{40}{\nm} thick Cr layer using EBL and PMMA resist.

\end{itemize}

The final step adding the Ti/Au layer was only performed in response to high noise levels during measurements and poor SNR, significantly improving device performance.
This leaves the finished embedded nanomagnet microstrip to be wirebonded to a suitable PCB to be incorporated into the MRFM system. In case of multiple microstrips on a chip unit (which is preferable to increase yield), dicing of the chip is required beforehand. The final microstrip chip has a dimension of \SI{5}{\mm} x \SI{1.5}{\mm}. Fixation to the supporting PCB is achieved by using dried PMMA 950K as glue.

 %\newpage

\section{Adiabaticity Plots}

The performance characteristics of the two microstrip generations is compared in Fig.~\ref{fig:AdiaAppendix}. Spin signal is measured in dependence of RF pulse strength (shown here as AWG Amplitude) and frequency deviation. A considerable increase in spin signal is observed for the planar microstrip design in comparison to the stacked microstrip design.

\begin{figure}[!t]
\includegraphics[scale = 1]{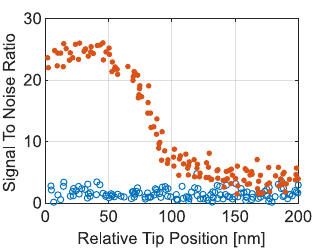}% size for 2column
\caption{SNR data corresponding to Fig.~3(c) showing maximum values of 25 at \SI{60}{\second} integration time, demonstrating excellent noise characteristics for the embedded nanomagnet design.}
\label{SNRAppendix}
\end{figure}

\section{SNR Measurements for Resolution Determination Measurement}

As a supplement for Fig.~3(c), the increase in SNR is shown for the corresponding measurement.

% \section{Appendix Deposit}%Fabrication of Embedded Nanomagnet Devices}

\newpage
~
\newpage
% \bibliography{references}
%\bibliography{references_zoteroRP}
%\input{references.bbl}
%
\end{document}